\documentclass[onecolumn]{revtex4}%
\usepackage{amssymb}
\usepackage{amsmath}
\usepackage{epsfig}
\usepackage{amsfonts}
\usepackage{graphicx}%
\begin{document}
\title{Analytical study of a gas of gluonic quasiparticles at high temperature:
effective mass, pressure and trace anomaly }
\author{Francesco Giacosa}
\affiliation{Institute for Theoretical Physics, Johann Wolfgang Goethe University,
Max-von-Laue-Str.\ 1, D--60438 Frankfurt am Main, Germany}

\begin{abstract}
The thermodynamical properties of a pure Yang-Mills theory $SU(N)$ is
described by a gas of gluonic quasiparticles with temperature dependent mass
$m(T)$ and a bag function $B(T).$ The analytic behavior of $m(T)$ and the
pressure $p$ in the temperature range $2.5$-$5T_{c}$ are derived and
constraints on the parameters defining $B(T)$ are discussed. The trace anomaly
$\theta=\rho-3p$ is evaluated in the high $T$ domain: it is dominated by a
quadratic behavior $\theta=nKT^{2},$ where $n=2(N^{2}-1)$ is the number of
degrees of freedom and $K$ is an integration constant which does not depend on
the bag function $B(T)$. The quadratic rise of $\theta$ is in good agreement
with recent lattice simulations.

\end{abstract}

\maketitle

\section{Introduction and Summary}

The concept of quasiparticle is a valuable tool toward an effective
description of complicated interactions. An important example is that of
Yang-Mills (YM) $SU(N)$ theories at nonzero temperature $T$, where their
intrinsic nonperturbative nature renders the applicability of a perturbative
treatment limited \cite{linde}. Descriptions of the YM system at $T>0$ via
gluonic quasiparticles have been performed in a variety of approaches, e.g.
Refs. \cite{dirk,weise,fabien,levai,drago,others,others2} and refs. therein.
Different \textit{Ans\"{a}tze} have been tested and the outcoming
thermodynamical quantities, such as energy density and pressure, have been
compared to lattice results \cite{karsch,boyd,cheng}. An appropriate, but at
the same time analytically tractable, description of a gas of gluons is also a
necessary step toward the understanding of the quark gluon plasma, see for
instance Ref. \cite{dirkrev} for a review.

Besides the temperature dependent gluonic mass $m=m(T)$, also a temperature
dependent bag energy $B=B(T)$ has been widely used to describe nonperturbative
properties of YM theories, such as the trace anomaly and the gluon condensate.
With these two basic ingredients the energy density and the pressure read (see
Refs. \cite{weise,fabien} and refs. therein):%
\begin{equation}
\rho=\rho_{p}+B(T)\text{ , }p=p_{p}-B(T)\text{ }, \label{rhop}%
\end{equation}
where the suffix `$p$' denotes the quasiparticle part:
\begin{equation}
\rho_{p}=n\int_{k}\frac{\sqrt{k^{2}+m^{2}(T)}}{\exp\left[  \frac{\sqrt
{k^{2}+m^{2}(T)}}{T}\right]  -1},%
\begin{array}
[c]{c}
\end{array}
\text{ }p_{p}=-Tn\int_{k}\log\left[  1-\exp\left[  -\frac{\sqrt{k^{2}%
+m^{2}(T)}}{T}\right]  \right]  \text{ }, \label{rhop2}%
\end{equation}
where $\int_{k}=\int$ $\frac{d^{3}k}{(2\pi)^{3}}=\int_{0}^{\infty}\frac
{k^{2}dk}{2\pi^{2}}$ and $n$ represents the degeneracy of the particle
spectrum. Restricting to perturbative degrees of freedom only, one has
$n=2(N^{2}-1)$ in the case of a $SU(N)$ YM theory.

In this article we study the high $T$ properties of the gluonic gas expressed
in Eqs. (\ref{rhop}) and (\ref{rhop2}). On the practical side, we are
interested in the behavior in the range between -say- $2.5T_{c}$ and $5T_{c}$,
where $T_{c}$ is the critical temperature for deconfinement, above which
gluonic quasiparticles are the relevant degrees of freedom. In this energy
range lattice data for the energy density and pressure show a plateau
\cite{karsch,boyd,cheng}, which is compatible with a linear increase of the
quasiparticle mass with the temperature $T,$ $m\propto T.$ Because of this
linear increase the Stefan-Boltzmann limit is not reached, but a saturation at
lower values is obtained. This situation persists as long as logarithmic
corrections due to the running coupling of QCD are neglected. In fact, the
correct perturbative behavior is $m\propto T/\sqrt{\log(T/\Lambda_{QCD})},$
which implies a slow approach of the pressure and energy density to the
expected Stefan-Boltzmann values, see the lattice simulation in Ref.
\cite{fodor}. In the present work we do not include these logarithmic
corrections. This simplification is applicable in the temperature range
$2.5$-$5T_{c},$ which is high when compared to the critical temperature, but
not high enough for the logarithmic terms to become relevant.

Strong constraints on the gas of quasiparticles can be derived by imposing
that the system fulfills thermodynamical self-consistency
\cite{dirk,weise,fabien,levai,drago}, which is a consequence of the first
principle of thermodynamics:
\begin{equation}
\rho=T\frac{dp}{dT}-p\text{ .}\label{tdsc}%
\end{equation}
The bag constant $B(T)$ is assumed to have the following behavior:%
\begin{equation}
B(T)=B_{NP}(T)+B_{P}(T)\text{ }\label{bagconst}%
\end{equation}%
\begin{equation}
\text{with }B_{NP}(T)=ncT^{\alpha}\text{ and }B_{P}(T)=n\delta T^{4}\text{ for
}2.5T_{c}\lesssim T\lesssim5T_{c}\text{ .}%
\end{equation}

The term $B_{P}(T)=n\delta T^{4}$ is included in order to recover the expected
results of perturbation theory in the high temperature regime, where the
effective gluon mass grows linearly with $T$ (up to the previously mentioned
logarithmic corrections, which are not considered here.)

The term $B_{NP}(T)=ncT^{\alpha}$, where $\alpha$ is a real number smaller
than $4$ and $c$ is a constant with the dimension of [Energy]$^{4-\alpha}$,
describes the `nonperturbative' bag contribution relevant above the phase
transition \cite{footnote}. At the present stage the only and general
requirement about the function $B_{NP}(T)$ is that it is dominated by a
power-like term $T^{\alpha}$ in the high $T$ region. It is the aim of this
work to constrain the value of $\alpha$ and $c$ by using mathematical
considerations and lattice results.

As a last remark we stress that the bag function $B(T)=B_{NP}(T)+B_{P}(T)$ is
proportional to the number of degrees of freedom $n=2(N_{c}^{2}-1),$ in
agreement with general large $N$ scaling arguments \cite{diakonov}.

The temperature dependent mass $m(T)$ can be analytically evaluated at high
$T$ (details are in Sec. 2) and takes the form ($\alpha\neq2$):
\begin{equation}
m(T)=T\sqrt{\frac{4\pi^{2}}{D(a_{0})}\frac{\alpha c}{2-\alpha}T^{\alpha
-4}+k\frac{\Lambda^{2}}{T^{2}}-\frac{8\pi^{2}\delta}{D(a_{0})}}\text{ ,}
\label{massintro}%
\end{equation}
where $D(a_{0})$ is a positive real number which shall be specified later on,
$\alpha,c,\delta$ are the already introduced parameters defining $B(T)$, $n$
is the number of degrees of freedom, $\Lambda$ is the Yang-Mills scale and
finally $k$ is an integration constant related to the differential equation
(\ref{tdsc}): $k$ is not determined by the choice of $B(T)$ but is a further
parameter entering in the model.

Mathematically and physically based considerations about Eq. (\ref{massintro})
will lead us to establish nontrivial relations between the parameters:

$\bullet$ The parameter $\delta$ parametrizes the linear increase of $m$ with
$T.$ As evident from Eq. (\ref{massintro}), $\delta\leq0$ in order to avoid an
imaginary mass at high $T$. Moreover, a careful study of the equations will
lead us to establish also an upper limit on its absolute value: $\left\vert
\delta\right\vert \leq0.0151$.

$\bullet$ Many approaches show that the nonperturbative bag function $B_{NP}$
is a positive number \cite{diakonov,bag}. In the present framework it means
that $c>0.$ Moreover, the contribution of nonperturbative physics to the
effective gluon mass should be positive, a property which also assures that no
instability emerges at low $T.$ Thus, it is a general physical requirement
that the nonperturbative contribution to $m^{2}(T)$ given by $\frac{4\pi^{2}%
}{D(a_{0})}\frac{\alpha c}{2-\alpha}T^{\alpha-2}$ in Eq. (\ref{massintro}) is
also positive. This, in turn, implies a consistent limitation on the choice of
the parameter $\alpha$: $0\leq\alpha<2$. Note, for the very same requirement
we are led to conclude that the integration constant $k$ is positive.

$\bullet$ The case $\alpha=2$ is somewhat peculiar because of the emergence of
logarithms in the solution, see details in Sec. 2.4. However, it is also
unfavoured because of similar arguments.

As a next step of this work we turn to the explicit expression for the
pressure $p$ in the high $T$ limit, see Sec. 3 for details. We shall find that
$p$ is expressed by the sum of three terms ($\alpha\neq2$):
\begin{equation}
p=-nc\frac{2}{2-\alpha}T^{\alpha}-\frac{n}{4\pi^{2}}D(a_{0})k\Lambda^{2}%
T^{2}\text{ }+\left(  \overline{p}_{p}(a_{0})-n\delta\right)  T^{4}\text{ .}
\label{pintro}%
\end{equation}
The first, negative term scale as $T^{\alpha}$ (just as $B_{NP}$); the second,
also negative, term scales as $T^{2}$ (in agreement with the phenomenological
argument of Ref. \cite{pisarski}) and is proportional to the constant $k$; the
third, positive term goes as $T^{4}$, but the coefficient is slightly smaller
than the Stefan-Boltzmann limit, see later on for details. The overall
pressure scale as $n\propto N^{2}$ in agreement with large $N$ scaling arguments.

The final and main subject of the present work is the study of the trace
anomaly at high $T$ (Sec. 4). The trace anomaly $\theta$, defined as%
\begin{equation}
\theta=\rho-3p=4B+\rho_{p}-3p_{p}\text{ ,}%
\end{equation}
has been in the center of a vivid debate in the last years (see Refs.
\cite{pisarski,miller,lingrow,zw,salcedo,laine} and refs. therein). Strict
dilatation invariance would imply that $\theta$ vanishes in a dilatation
invariant theory, such as a gas of photons. In a YM theory this symmetry is
broken by quantum effects and $\theta$ does not vanish: this is the so-called
trace anomaly. We aim to show that, in the context of a gas of quasiparticle
with the general form of the bag constant $B(T)$ given in Eq. (\ref{bagconst}%
), the following high $T$ behavior holds:%
\begin{equation}
\theta=\rho-3p=nCT^{\alpha}+nKT^{2}\text{ for }T\gtrsim2T_{c}\text{ ,}
\label{thetaintro}%
\end{equation}
where $C=2c\frac{4-\alpha}{2-\alpha}$ is a constant determined by the
`nonperturbative' parameters of the model $\alpha,c$ (i.e., those parameters
which define the nonperturbative bag function $B_{NP}(T)=ncT^{\alpha}$), and
$K=\frac{k\Lambda^{2}D(a_{0})}{2\pi^{2}}$ is proportional to the previously
introduced integration constant $k$. The main result is that the trace anomaly
$\theta$ can be decomposed in a term which behaves as the `nonperturbative'
contribution to the bag constant $B_{NP}(T)=ncT^{\alpha}$, and a term which
goes as $T^{2}$. Restricting to the favoured interval $0\leq\alpha<2,$ one is
led to conclude that the quadratic rise dominates at $T$ large enough.
Remarkably, the $T^{2}$ rise of $\theta$ is a general property, which is
independent on the nonperturbative bag constant $B_{NP}$.

It is indeed remarkable that a quadratic rise of the trace anomaly,
$\theta\simeq aT^{2}$, is found in Ref. \cite{pisarski}, where an analysis of
the lattice data of Ref. \cite{boyd} has been performed. Later on, this
quadratic rise has been confirmed in recent lattice works \cite{cheng,panero}.
In particular, in the lattice study of Ref. \cite{panero} the trace anomaly
has been investigated for various pure Yang-Mills theories $SU(N)$,
$N=3,...,8$. The behavior $\theta\propto nT^{2}$ (with a direct
proportionality to the degeneracy number $n=2(N^{2}-1)$) is indeed found for
$N=3,...,8$ in a range between $2$ and $5T_{c}$. Thus, the result of Eq.
(\ref{thetaintro}) may explain in a natural way the emergence of such
quadratic behavior of $\theta$ at high $T$ .

In the following sections we present the detailed derivations of the outlined
results: in Sec. 2 and 3 we derive the expressions for $m(T)$ in Eq.
(\ref{massintro}) and for the pressure $p$ in Eq. (\ref{pintro}). In Sec. 4 we
present the calculation leading to the expression of the trace anomaly
$\theta(T)$ in Eq. (\ref{thetaintro}). Finally, in Sec. 5, we briefly outline
our conclusions.

\section{Temperature-dependent quasiparticle mass $m(T)$}

\subsection{Differential equation for $m(T)$}

In order to obtain the differential equation for the quasiparticle mass
$m=m(T)$, we plug the expressions of Eqs. (\ref{rhop}) and (\ref{rhop2}) into
the the thermodynamical self-consistency relation $\rho=T\frac{dp}{dT}-p$:
\begin{equation}
\frac{dB}{dT}=-nI(m)\frac{dm^{2}}{dT},%
\begin{array}
[c]{c}
\end{array}
\text{ }I(m)=\int_{k}\frac{1}{2\sqrt{k^{2}+m^{2}}}\frac{1}{\exp\left[
\frac{\sqrt{k^{2}+m^{2}(T)}}{T}\right]  -1}\text{ ,} \label{eq1}%
\end{equation}
where the bag function $B(T)$ is given in Eq. (\ref{bagconst}).

\subsection{Use of dimensionless functions}

It is convenient to rewrite the equations by using dimensionless quantities.
To this end we introduce the dimensionless temperature
\begin{equation}
\lambda=\frac{T}{\Lambda}\text{ ,} \label{lamda}%
\end{equation}
where $\Lambda$ is the Yang-Mills scale, which is of the same order of the
critical temperature $T_{c},$ $\Lambda\sim T_{c}.$

The dimensionless particle contribution to the energy density and pressure%
\begin{equation}
\overline{\rho}_{p}=\frac{\rho_{p}}{T^{4}}\text{ , }\overline{p}_{p}%
=\frac{p_{p}}{T^{4}}%
\end{equation}
read%
\begin{align}
\overline{\rho}_{p}  &  =\frac{n}{2\pi^{2}}\int_{0}^{\infty}dx\frac{x^{2}%
\sqrt{x^{2}+a^{2}}}{e^{\sqrt{x^{2}+a^{2}}}-1}\text{ ,}\\
\overline{p}_{p}  &  =-\frac{n}{2\pi^{2}}\int_{0}^{\infty}dx\left(  x^{2}%
\ln\left(  1-e^{-\sqrt{x^{2}+a^{2}}}\right)  \right)  \text{ .} \label{pbarp}%
\end{align}
The function $a=a(\lambda)$ is the `dimensionless mass':
\begin{equation}
a=a(\lambda)=\frac{m(T)}{T}=\frac{m(\lambda\Lambda)}{\lambda\Lambda}\text{ .}
\label{a}%
\end{equation}
We also define the dimensionless constant $\gamma$ as%
\begin{equation}
\gamma=c\Lambda^{\alpha-4}\text{ .}%
\end{equation}
In this way the ground-state dimensionless energy density $\overline{\rho
}_{gs}$ and pressure $\overline{p}_{gs}$
\begin{equation}
\overline{\rho}_{gs}=-\overline{p}_{gs}=\frac{\rho_{gs}}{T^{4}}=\frac
{B(T)}{T^{4}}%
\end{equation}
read explicitly (see Eq. (\ref{bagconst}))%
\begin{equation}
\overline{\rho}_{gs}=-\overline{p}_{gs}=n\gamma\lambda^{\alpha-4}%
+n\delta\text{ .}%
\end{equation}
The full dimensionless energy-density and pressure read%
\begin{equation}
\overline{\rho}=\overline{\rho}_{p}+\overline{\rho}_{gs}\text{ , }\overline
{p}=\overline{p}_{p}+\overline{p}_{gs}\text{ .}%
\end{equation}
The thermodynamical self-consistency of Eq. (\ref{tdsc}) can be rewritten in
terms of the reduced energy density and pressure as%
\begin{equation}
\overline{\rho}-3\overline{p}=\lambda\frac{d\overline{p}}{d\lambda}\text{ .}
\label{tdscdimles}%
\end{equation}

In terms of the dimensionless quantities he differential Eq. (\ref{eq1}) takes
the form:%

\begin{equation}
n\alpha\gamma\lambda^{\alpha-4}+4n\delta=-\frac{n}{2\pi^{2}}\frac
{d(a^{2}\lambda^{2})}{d\lambda}\frac{D(a)}{2\lambda}\text{ ,} \label{diffeq}%
\end{equation}
where the integral $D(a)$ is given by%
\begin{equation}
D(a)=\int_{0}^{\infty}dx\frac{x^{2}}{\sqrt{x^{2}+a^{2}}}\frac{1}%
{e^{\sqrt{x^{2}+a^{2}}}-1}\text{ .} \label{da}%
\end{equation}
Note, the dependence on the degeneracy number $n=2(N^{2}-1)$ factorizes, so
that the equation for $a(\lambda)$ is independent on the number of colors $N$.
This is in agreement with the general expectation of large $N$ scaling,
according to which the effective gluon mass $m(T)$ scales as $N^{0}.$

\subsection{Constraints on the parameter $\delta$}

In the limit of large $\lambda$ (i.e., large $T)$ one has $a(\lambda
>>\lambda_{c})\rightarrow a_{0}.$ In this way, besides the logarithmic
corrections, the effective mass exhibits a linear growth $m=a_{0}T,$ in
agreement with the expectation of perturbation theory \cite{kapusta} and with
high $T$ effective approaches, e.g. Ref. \cite{blaizot}. In the present
phenomenological approach the numerical value of $a_{0}$ is related to the
parameter $\delta$ by studying the asymptotic behavior of Eq. (\ref{diffeq}):
\begin{equation}
\delta=-\frac{1}{8\pi^{2}}a_{0}^{2}D(a_{0})\text{ .}%
\end{equation}
In Fig. 1, left panel, the quantity $\delta$ is plotted as function of
$a_{0}.$ The two properties mentioned in the Introduction can be easily
proven: $\delta\leq0$ and $\left\vert \delta\right\vert \leq\max$ $\left(
\frac{1}{8\pi^{2}}a_{0}^{2}D(a_{0})\right)  =0.0151.$%

\begin{figure}
[ptb]
\begin{center}
\includegraphics[
height=2.2978in,
width=6.3815in
]%
{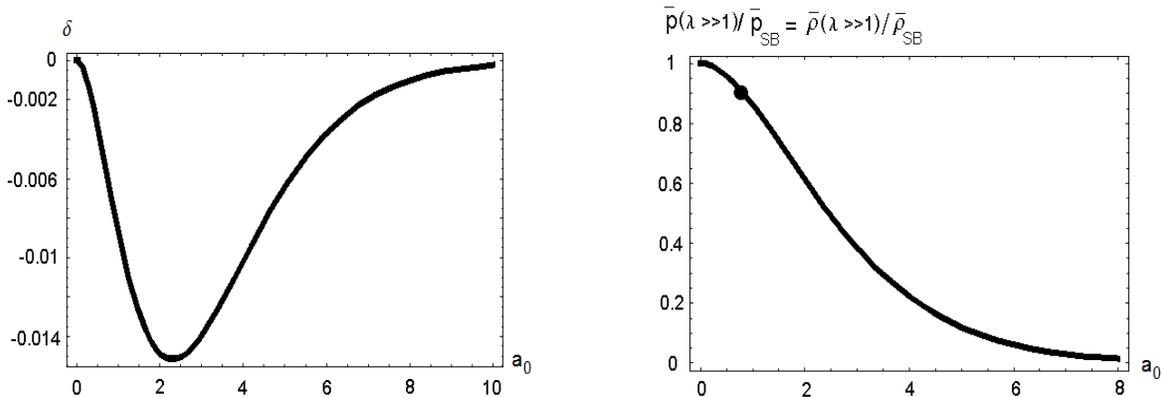}%
\caption{Left panel: the parameter $\delta$ is plotted as function of the
asymptotic value $a(\lambda\rightarrow\infty)=a_{0}.$ Right panel: the ratio
$(\overline{p}/\overline{p}_{SB})_{\lambda>>1}=(\overline{\rho}/\overline
{\rho}_{SB})_{\lambda>>1}$ is plotted as function of $a_{0}.$ The dot
corresponds to $\overline{\rho}/\overline{\rho}_{SB}=0.9$ and $a_{0}=0.83;$
this is the saturation value obtained in lattice simulations \cite{panero}.}%
\label{fig1}%
\end{center}
\end{figure}

Various lattice simulations of Yang-Mills system \cite{boyd,cheng} show that
the Stefan-Boltzmann limit of the energy density and the pressure
\begin{equation}
\overline{\rho}_{SB}=n\frac{\pi^{2}}{30}\text{ , }\overline{p}_{SB}=n\frac
{\pi^{2}}{90}%
\end{equation}
is not reached at $5T_{c}$. On the contrary, a saturation at a lower value of
about $90\%$ of the Stefan-Boltzmann limit is observed. Such a saturation is
obtained in the present quasiparticle approach by a nonzero value of $a_{0}$
(i.e., a nonzero value of $\delta).$ For high $\lambda$ the function
$\overline{p}(\lambda)/\overline{p}_{SB}$ approaches the asymptotic value
$(\overline{p}_{p}(a_{0})-n\delta)/\overline{p}_{SB}$ (see Sec. III). In Fig.
1, right panel, the quantity $\left(  \overline{p}(\lambda)/\overline{p}%
_{SB}\right)  _{\lambda>>1}=(\overline{\rho}(\lambda)/\overline{\rho}%
_{SB})_{\lambda>>1}$ is plotted as a function of $a_{0}.$ In order that at
high $\lambda$ the ratio $\overline{\rho}(\lambda)/\overline{\rho}_{SB}%
\simeq0.9$ holds, the value $a_{0}\simeq0.83$ is required. This, in turn,
implies that $\delta\simeq-0.0070.$ Note, similar values for $a_{0}$ have been
obtained in Refs. \cite{fabien,levai}.

As explained in the Introduction, the Stefan-Boltzmann limit for the energy
density and the pressure is reached at much higher temperatures \cite{fodor},
at which the logarithmic decrease of $a(\lambda)$ becomes relevant.

\subsection{Analytical solution $a(\lambda)$ in the large $\lambda$ domain
(i.e., $m(T)$ in the large $T$ domain)}

An analytical solution of Eq. (\ref{diffeq}) can be obtained in the limit of
large $\lambda$ by approximating the function $D(a)$ by its asymptotic values
$D(a_{0})$. In this limit Eq. (\ref{diffeq}) can be easily solved and one
obtains for $a^{2}(\lambda)$ ($\alpha\neq2$):%
\begin{equation}
a^{2}(\lambda)=\frac{4\pi^{2}}{D(a_{0})}\frac{\alpha\gamma}{2-\alpha}%
\lambda^{\alpha-4}+\frac{k}{\lambda^{2}}+a_{0}^{2}\text{ ,} \label{ylamda}%
\end{equation}
%

\begin{figure}
[ptb]
\begin{center}
\includegraphics[
height=2.6005in,
width=4.0162in
]%
{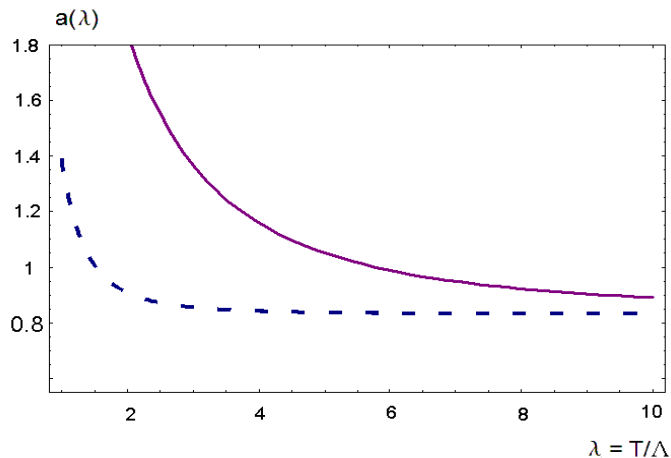}%
\caption{Comparison of the analytic, approximate expression of Eq.
(\ref{ylamda}) (solid line) with the full solution of the differential Eq.
(\ref{diffeq}) (dashed line) for a particular numerical case: $\alpha=1,$
$\gamma=0.034,$ $k=10.$ }%
\label{fig2}%
\end{center}
\end{figure}

\bigskip

In Fig. 2 the analytical expression (\ref{ylamda}) is compared to the
numerical result of the differential Eq. (\ref{diffeq}) for a particular
numerical choice; the numerical solution takes into account the full
dependence on the function $D(a),$ which is ignored in the analytical one.
However, the qualitative agreement, which becomes better and better by
increasing $\lambda$, is visible.

The analytical expression of the function $m^{2}(T)$ is easily obtained by
making use of Eqs. (\ref{lamda}), (\ref{a}) and (\ref{ylamda}):%
\begin{equation}
m^{2}(T)=\frac{4\pi^{2}}{D(a_{0})}\frac{\alpha c}{2-\alpha}T^{\alpha
-2}+k\Lambda^{2}+a_{0}^{2}T^{2}\text{ .} \label{m2}%
\end{equation}
As a result of the obtained expression for $m(T)$ we can discuss the
constraints on the parameters $c$ and $\alpha$:

(i) The squared mass constitutes of three terms. The first term in Eq.
(\ref{m2}) depends on the parameters $c$ and $\alpha,$ which define the
nonperturbative bag pressure $B_{NP}=cT^{\alpha}.$ The second contribution in
Eq. (\ref{m2}), $k\Lambda^{2}$, is constant and is directly proportional to
the integration constant $k.$ The last term in Eq. (\ref{m2}) dominates at
high $T$, implying the linear behavior $m(T)\simeq a_{0}T$, as predicted by
perturbative calculations and effective theories of QCD. As already
anticipated in\ Sec. 1, the natural requirements that $B_{NP}=cT^{\alpha}$ and
the corresponding contribution to $m^{2}(T),$ given by $\frac{4\pi^{2}%
}{D(a_{0})}\frac{\alpha c}{2-\alpha}T^{\alpha-2},$ are positive numbers
implies that: $\gamma>0,$ $0\leq\alpha<2.$ Similarly, $k>0$.

(ii) In the limit $\alpha=0$ the first term in Eq. (\ref{m2}) vanishes and the
second, constant contribution dominates: this situation corresponds to the
simple approximation with a constant gluon mass. This possibility has however
been ruled out by a precise comparison with lattice data \cite{fabien}. More
generally, when $\alpha$ is small, the first term in Eq. (\ref{m2}) is also
negligible (unless the parameter $c$ is anomalously large); at low $T$ only
the second constant term survives. On the contrary, when $\alpha\simeq2$ the
first term is very large, unless the parameter $c$ is very small. We thus
conclude that $\alpha$ should be not to close to the boundaries $0$ and $2,$
but somewhere in between.

(iii) In the case $\alpha=2$ a slightly different solution is obtained:%
\begin{equation}
a^{2}(\lambda)=-\frac{8\pi^{2}\gamma}{D(a_{0})}\frac{\log\lambda}{\lambda^{2}%
}+\frac{k}{\lambda^{2}}+a_{0}^{2}\text{ ,}%
\end{equation}
where an additional logarithm arises. The mass contribution of the
nonperturbative first term is negative for $\gamma>0$ (that is for the here
considered choice $c>0$). Thus, also the case $\alpha=2$ is regarded as
phenomenologically unfavoured.

(iv) In order to include the logarithmic corrections in the very high $T$
domain one should modify the perturbative bag pressure $B_{P}(T,)$ in such a
way that the perturbative mass behavior $m\propto T/\sqrt{\log T/\Lambda}$
holds. One obtains the constrain $T^{-3}dB_{P}/dT\propto\left(  2\ln
^{-1}\lambda-2\ln^{-2}\lambda\right)  ,$ thus leading to more complicated
expressions involving logarithms. A detailed study of this subject represent
an interesting outlook. Although the formulas will be more involved, a link
with studies of Ref. \cite{others}, in which the starting point is the
perturbative behavior at very large temperature, can be driven.

\section{Pressure in the high $T$ domain}

We turn to the explicit expression of the pressure for large $T$. To this end
we expand Eq. (\ref{pbarp}) around the asymptotic value $a_{0}^{2}$:%
\begin{equation}
\overline{p}_{p}(a)=\overline{p}_{p}(a_{0})+\left(  \frac{d\overline{p}%
_{p}(a)}{da}\right)  _{a_{0}}(a-a_{0})+...
\end{equation}
Using the equality $\left(  \frac{d\overline{p}_{p}(a)}{da}\right)  _{a_{0}%
}=-\frac{n}{2\pi^{2}}a_{0}D(a_{0})$ and approximating $a_{0}(a-a_{0}%
)=a^{2}-a_{0}^{2}$ (valid at the considered order) one gets%
\begin{equation}
\overline{p}_{p}=\overline{p}_{p}(a_{0})-n\frac{\alpha\gamma}{2-\alpha}%
\lambda^{\alpha-4}-\frac{n}{4\pi^{2}}D(a_{0})\frac{k}{\lambda^{2}}\text{ .}%
\end{equation}
The full dimensionless pressure $\overline{p}=$ $\overline{p}_{p}+\overline
{p}_{gs}$ reads at high $\lambda$:%
\begin{equation}
\overline{p}=-n\gamma\frac{2}{2-\alpha}\lambda^{\alpha-4}-\frac{n}{4\pi^{2}%
}D(a_{0})\frac{k}{\lambda^{2}}+\left(  \overline{p}_{p}(a_{0})-n\delta\right)
\text{ .}%
\end{equation}
By multiplying by $T^{4}$ we find the pressure $p$ for large $T$:%
\begin{equation}
p=-nc\frac{2}{2-\alpha}T^{\alpha}-\frac{n}{4\pi^{2}}D(a_{0})k\Lambda^{2}%
T^{2}\text{ }+\left(  \overline{p}_{p}(a_{0})-n\delta\right)  T^{4}\text{.}
\label{phight}%
\end{equation}
We thus have also decomposed the pressure into three contributions: the first
term in\ Eq. (\ref{phight}) scales exactly as the bag function $B_{NP}.$ The
second, negative term in Eq. (\ref{phight}) scales as $T^{2}$ and is
proportional to the integration constant $k.$ Note, a similar quadratic
contribution to the pressure has been postulated in Ref. \cite{pisarski}.
There is, however, an important point to stress: here we have shown that there
is is no need to introduce at hand a quadratic contribution to the pressure
from the very beginning. \emph{The quadratic contribution naturally emerges as
the result of the equation, independently on the choice of the bag function
}$B_{NP}(T).$ Finally, the last term in Eq. (\ref{phight}) describes the high
$T$ asymptotic limit, which differs from the Stefan-Boltzmann value as
depicted in the right panel of Fig. 1. For other works on the pressure in the
high $T$ domain see Refs. \cite{zw2,laine2} and refs. therein. For a direct
comparison with lattice data see Fig. 3.

\section{Trace anomaly}

The interaction measure%
\begin{equation}
\Delta=\frac{\theta}{T^{4}}=\overline{\rho}-3\overline{p}%
\end{equation}
is evaluated by making use of the thermodynamical self-consistency of Eq.
(\ref{tdscdimles}):%
\begin{equation}
\Delta=\overline{\rho}-3\overline{p}=\lambda\frac{d\overline{p}}{d\lambda
}=\lambda\frac{d\overline{p}_{p}}{d\lambda}+\lambda\frac{d\overline{p}_{gs}%
}{d\lambda}\text{ ,}%
\end{equation}
where in the last step the dimensionless pressure has been decomposed into its
particle and ground-state contributions.

The ground-state contribution is easily evaluated:%
\begin{equation}
\lambda\frac{d\overline{p}_{gs}}{d\lambda}=n\gamma(4-\alpha)\lambda^{\alpha
-4}.
\end{equation}
The calculation of the particle contribution to the interaction measure goes
via two steps. First, we rewrite it by making use of Eqs. (\ref{pbarp}):%
\begin{equation}
\lambda\frac{d\overline{p}_{p}}{d\lambda}=-\frac{n}{2\pi^{2}}\lambda
D(a)a\frac{da}{d\lambda}=-\frac{n}{4\pi^{2}}\lambda D(a)\frac{da^{2}}%
{d\lambda}\text{ .}%
\end{equation}
As a second step, in the large $\lambda$ domain one can replace $D(a)$ with
the asymptotic value $D(a_{0})$ and then evaluates the derivative
$\frac{da^{2}}{d\lambda}$ by using Eq. (\ref{ylamda}):%
\begin{equation}
\lambda\frac{d\overline{p}_{p}}{d\lambda}=n\alpha\gamma\frac{4-\alpha
}{2-\alpha}\lambda^{\alpha-4}+n\frac{D(a_{0})}{2\pi^{2}}\frac{k}{\lambda^{2}%
}\text{ .}%
\end{equation}
By putting the results together one finds for $\lambda\gtrsim2\lambda_{c}$
($\alpha\neq2$)%
\begin{equation}
\Delta=\overline{\rho}-3\overline{p}=2n\gamma\frac{4-\alpha}{2-\alpha}%
\lambda^{\alpha-4}+n\frac{D(a_{0})}{2\pi^{2}}\frac{k}{\lambda^{2}}\text{ ,}%
\end{equation}
which consists of two terms: a term which scales as the ground-state
contribution $\overline{\rho}_{gs}$, and a term which scales as $\lambda
^{-2}.$

By multiplying $\Delta$ by $T^{4}$ one obtains the trace anomaly $\theta$ as
function of $T$ ($\alpha\neq2$):%
\begin{equation}
\theta=2nc\frac{4-\alpha}{2-\alpha}T^{\alpha}+n\frac{D(a_{0})}{2\pi^{2}%
}k\Lambda^{2}T^{2}\text{ ,} \label{thetafin}%
\end{equation}
which is Eq. (\ref{thetaintro}) discussed in the Introduction: the first term
depends on $c$ and $\alpha$ (i.e., the parameters which define $B_{NP}(T)$),
while the second term describes a quadratic rise of $\theta$, is proportional
to the integration constant $k$ and is independent on the bag function $B$.
The very same term proportional to $k$ was responsible for a constant
contribution to the effective gluon mass, see Eq. (\ref{m2}). In the favoured
range $0\leq\alpha<2$ the rise $T^{\alpha}$ is realized for small $T,$ while
the quadratic rise of $\theta$ dominates for large enough temperature. The
temperature at which this change happens depends on the particular numerical
values of the parameters, and cannot be determined by analytical considerations.

It is however possible to use some lattice results about the trace anomaly in
order to constrain the numerical values of the parameters of the model. The
quantity $\theta$ scales as $3.3T_{c}^{2}/T^{2}$ for $T\gtrsim1.5$-$2T_{c}$
\cite{fabien}. Then, from Eq. (\ref{thetafin}) and $\Lambda\sim T_{c}$ it
follows that $k\simeq10.$ It is also possible to obtain a rough estimation of
the upper limit of the parameter $\gamma=c\Lambda^{4-\alpha}\sim
cT_{c}^{4-\alpha}.$ In fact, the quantity $\theta/T^{2}$ is, to a good
approximation, constant for $T\gtrsim2T_{c}$ \cite{pisarski}. This, in turn,
means that the first term in Eq. (\ref{thetafin}) is smaller than the second
term for $T\gtrsim2T_{c}$. Through simple algebra one obtains the upper limit
$\gamma\lesssim\frac{2-\alpha}{4-\alpha}\frac{D(a_{0})k}{2\pi^{2}}2^{1-\alpha
}$.

We now turn to a direct comparison of our theoretical curves with the lattice
results of Ref. \cite{boyd}. Since our theoretical functions depend on the
variable $\lambda=T/\Lambda,$ where $\Lambda\sim T_{c}$ but not exactly equal,
care is needed: it is first necessary to determine $\Lambda.$ To this end we
chose $\Lambda$ in such a way that the theoretical result for the interaction
measure $\Delta$ reproduces the lattice point at the highest simulated value
of $T/T_{c}=4.57$ (at which $\Delta=0.10$). One obtains the relation
$\Lambda=1.55T_{c}.$ In Fig. 3 the plot of the interaction measure (left
panel) and energy and pressure (right panel) are shown: it is visible that the
agreement is acceptable for $T\gtrsim2.5T_{c}$ and increases for increasing
$T.$ On the contrary the theoretical results for $T/T_{c}\lesssim2.5$ deviate
from the lattice simulations. This is expected because the present version of
the model cannot describe the physical properties close to the phase transition.

\bigskip%

\begin{figure}
[ptb]
\begin{center}
\includegraphics[
height=2.5019in,
width=6.6971in
]%
{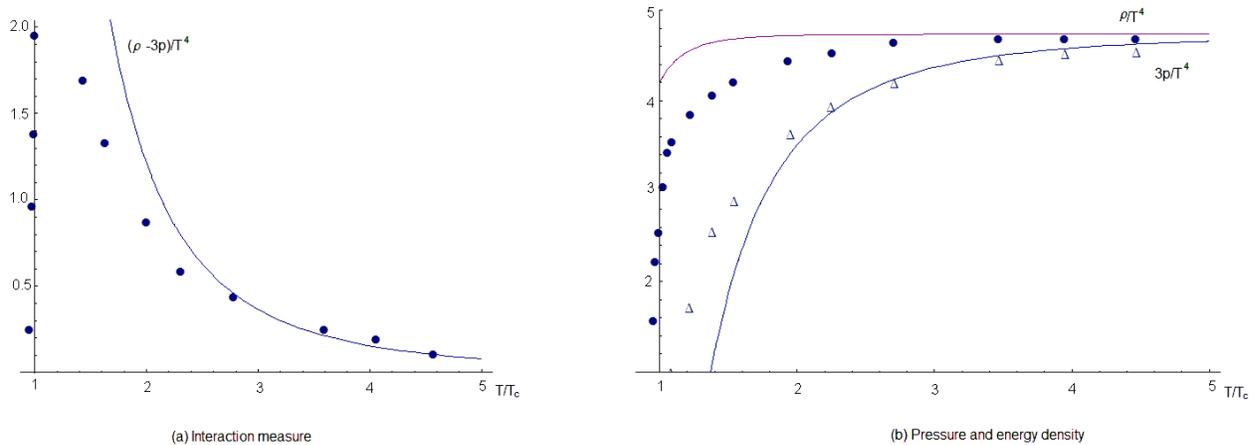}%
\caption{Comparison with the lattice results of Ref. \cite{boyd}. The
parameters $\alpha=1,$ $\gamma=0.034,$ $k=10$ are used. Left panel: the
interaction measure is plotted. The lattice point for $T/T_{c}=4.58$ has been
uused to determine $\Lambda=1.66T_{c}.$ Right panel: the quantities $\rho$
(the upper curve is the theoretical result, the dots the lattice points) and
$3p$ (the lower curve is the theoretical result, the triangles are the lattice
points) are shown. }%
\label{fignew}%
\end{center}
\end{figure}

Further comments are in order:

(i) For the here considered case $c>0$ (i.e. for $B_{NP}>0$) a positive
contribution of the first term to the trace anomaly is obtained for
$0\leq\alpha<2$ (see Eq. (\ref{thetafin})), which is the same interval
outlined previously. If, on the contrary, $2<\alpha<4$ one would have a
negative $\theta$ for high enough $T$, in disagreement with all present
simulations. This represents a further confirmation of the outlined range of
$\alpha$.

(ii) The case $\alpha=2$ leads to a slightly modified form:%
\begin{equation}
\theta=4nT^{2}\left(  c-c\log\left(  \frac{T}{\Lambda}\right)  +\frac
{D(a_{0})}{8\pi^{2}}k\Lambda^{2}\right)  \text{ .}%
\end{equation}
A negative $\theta$ at high $T$ is realized. This fact is at first sight a
further argument against the choice $\alpha=2$. However, the appearance of the
logarithms in the solution implies that a full study of the present case is
only possible when the logarithmic corrections are taken into account.

(iii) The behavior of $\theta$ in the large $T$ domain as measured on the
lattice is still subject to an ongoing discussion. In the work of Ref.
\cite{miller}, also based on the lattice data of Ref. \cite{boyd}, it is found
that $\theta$ growths linearly (rather than quadratically) with $T$ :
$\theta=aT$ for $2.5T_{c}\lesssim T\lesssim5Tc,$ where $a\simeq1.5$ GeV$^{3}$
in the $SU(2)$ case and $a\simeq1.7$ GeV$^{3}$in the $SU(3)$ case. We also
refer to the similar results obtained in the older works of Refs.
\cite{Kallman,langfeld}. In Ref. \cite{bugaev} the linear rise has been
confirmed by studying the lattice data of Ref. \cite{cheng}. Our result
(\ref{thetaintro}) can indeed also account for an initial non-quadratic
behavior of $\theta$ (the linear one being realized for $\alpha=1,$ see
\cite{lingrow}), which persists as long as the quadratic term does not become dominant.

(iv) The linear rise of $\theta$ has been derived within the theoretical
framework described in Refs. \cite{ralfrev}, and further investigated in Refs.
\cite{lingrow,garfield}. The non-perturbative sector of $SU(2)$ or $SU(3)$ YM
theories is described by a composite, (adjoint-)scalar field $\phi$ in the
deconfined phase ($T>T_{c})$, which emerges as an `average' over calorons and
anticalorons (topological objects which correspond to instantons at nonzero
$T$ \cite{hist}) with trivial holonomy, see \cite{ralfrev} for a microscopic
derivation and \cite{garfield} for a macroscopic one. On a length scale
$l>\left\vert \phi\right\vert ^{-1}$ it is thermodynamically exhaustive to
consider only the average field $\phi$ and neglect the (unsolvable)
microscopic dynamics of all YM-field configurations, such as calorons and
monopoles. One can then build up an effective theory for YM-thermodynamics
valid for $T>T_{c}$, in which the scalar field $\phi$ acts as background field
coupled to the residual, perturbative gluons. On a phenomenological level it
contributes to the energy and pressure as a temperature-dependent bag constant
$B_{NP}=4\pi\Lambda^{3}T,$ i.e. with the parameters $\alpha=1$ and
$\gamma=4\pi$. Note, in the theoretical framework of Ref. \cite{ralfrev} the
constant $k$ was set to be very small, thus the quadratic rise starts to
dominate only at very high temperatures and for this reason does not affect
the phenomenology between $2.5T_{c}$ and $5T_{c}$. The linear growth with $T$
of the stress-energy tensor in the pure $SU(2)$ YM theory is obtained as
\cite{lingrow}: $\theta=\rho-3p\overset{T>2T_{c}}{\sim}24\pi\Lambda^{3}%
T\simeq(1.7$ GeV$^{3})T$. The coefficient $1.7$ GeV$^{3}$ is similar to $1.5$
GeV$^{3}$ found in Ref. \cite{miller}. A similar result holds in the $SU(3)$ case.

(v) A linear growth of the trace anomaly $\theta$ has also been obtained
within the theoretical approach described in Ref. \cite{zw}, in which a
Gribov-type dispersion relation is used. On the contrary, a quadratic rise of
$\theta$ is the result of Ref. \cite{salcedo}, in which a dimension-two
gluonic condensate is studied.

(vi) In the present work we concentrated on the high $T$ side. If we assume
that a power-like behavior $B_{NP}(T)=cT^{\alpha}$ is valid for lower
temperature, the quasi-particle gluon mass blows up at a critical temperature
$T_{c}$ (which depends on the numerical values of $c$ and $\alpha,$ e.g. in
Ref. \cite{ralfrev}). This fact may signalize a confinement/deconfinement
phase transition. For $T<T_{c}$ a quasi-particle description is no longer
possible and the system should be described by different degrees of freedom
(such as glueballs \cite{glueballs} and other nonperturbative states).

\section{Conclusions}

In this work we have performed an analytical study of the high temperature
properties of a gas of gluonic quasiparticles with a temperature-dependent bag
function. The expression of the quasiparticle mass $m(T)$, pressure and trace
anomaly have been derived analytically for large $T.$

The implications and constraints on the parameters of the bag function
$B(T)=B_{NP}(T)+B_{P}(T)$ have been discussed: for the nonperturbative
contribution $B_{NP}(T)=ncT^{\alpha}$ (with $c>0$) we have found the following
constraint on the parameter $\alpha$: $0\leq\alpha<2$. This result follows
from the requirements that the gluon mass $m(T)$ does not become imaginary for
decreasing $T$ and that the sign of the trace anomaly $\theta=\rho-3p$ at
large $T$ is positive, in agreement with lattice simulations. The behavior of
the trace anomaly $\theta=\rho-3p$ at high temperatures consists of two
contributions, $\theta$ $=n\#T^{\alpha}+n\#T^{2},$ with a first term which
goes as $T^{\alpha}$, just as the the nonperturbative bag function, and a
second term which goes as $T^{2}$ and dominates the high $T$ behavior. It is
remarkable that this quadratic contribution is general and does not depend on
the choice of the bag function. Such a quadratic behavior, and also the
expected scaling with the degeneracy number $n$, have been confirmed in the
lattice simulations of Refs. \cite{cheng,panero}.

A variety of improvements of the present approach represents an outlook for
the future: (i) Inclusion of the logarithmic corrections in order to make
contact with the very high $T$ behavior, in which the pressure and the energy
density slowly approach their Stefan-Boltzmann limit. (ii) On the
low-temperature side a fit should be performed in order to determine the
behavior of the nonperturbative bag pressure $B_{NP}(T)$ beyond the simple
power-law used in this work for analytical considerations. (iii) Calculation
of viscosities following Ref. \cite{redlich} can be performed. (iv) Inclusion
of further degrees of freedom: quarks for $T>T_{c}$ and confined states
(glueball and mesons) for $T<T_{c}.$

\bigskip

\textbf{Acknowledgments}: The author is deeply thankful to Ralf Hofmann for
long discussions about the problematic of Yang-Mills thermodynamics. Marco
Panero is acknowledged for useful comments about lattice results and Elina
Seel for the careful reading the manuscript.


\begin{thebibliography}{99}
\bibitem {footnote}The case $\alpha>4$ is surely exlcuded, because it would
imply that at large $T$ the energy density and the pressure would scale as
$T^{\alpha>4}$ and not as $T^{4}$.
\end{thebibliography}
\end{document}